\documentclass[
aps,prd,
showpacs,reprint,notitlepage,
amssymb,amsmath,amsfonts,mathrsfs,
nofootinbib,superscriptaddress,
floats,floatfix
]{revtex4-2}

\usepackage{graphicx}
\usepackage{xcolor}
\usepackage[colorlinks=true]{hyperref}
\hypersetup{citecolor=cyan,linkcolor=magenta}
\usepackage{multirow,array}
\DeclareMathAlphabet{\pazocal}{OMS}{zplm}{m}{n}
\usepackage{journals}
\newcommand{\nus}{\nu_{\text{s}}}
\newcommand{\nuk}{\tilde{\nu}}
\newcommand{\Oo}{\Omega_{\rm orb}}
\newcommand{\IC}{{\tilde \Theta}}

\definecolor{maroon}{cmyk}{0,0.87,0.68,0.32}

\begin{document}
\title{The Last Three Seconds: Packed Message delivered by Tides in Binary Neutron Star Mergers}

\date{\today}

\author{Hao-Jui Kuan}
\email{hao-jui.kuan@aei.mpg.de}
\affiliation{Max Planck Institute for Gravitational Physics (Albert Einstein Institute), 14476 Potsdam, Germany}
\affiliation{Theoretical Astrophysics, IAAT, University of T{\"u}bingen, T{\"u}bingen, D-72076, Germany}

\author{Kostas D. Kokkotas}
\affiliation{Theoretical Astrophysics, IAAT, University of T{\"u}bingen, T{\"u}bingen, D-72076, Germany}

\begin{abstract}
It is known that the leading-order tidal effects in gravitational waveforms can be quantified by tidal deformability, while higher order terms, e.g., harmonic overtones of Love number and dynamical tides, have not been well-investigated yet. The concept of a ``form factor'', which is different from while resembles the effective tidal deformability, for the tidal interactions between neutron stars in coalescing binaries is illustrated here. The form factor effectively incorporates the contribution of dynamical tides. The dependence of tidal form factor on tidal deformability, spins, and inclination angles is modeled and expressed in a closed form.
\end{abstract}
\maketitle

\section{Introduction}
The morphology of gravitational waveforms depends on almost all source parameters, and thus encodes a bunch of information about the radiating objects \cite{cutl93,blan95,jara96,kokk01,arun09,read09,damo12,prat20} (see \cite{diet21} for a recent review). However, a satisfactory knowledge of source parameters can only be acquired if the systematic bias can be well-controlled \cite{fava14,wade14}.
Among other effects to be better modeled, tidal dephasing in waveforms of binary neutron stars (BNS; we refer binaries to coalescing ones in what follows) is resulting from the deposition of orbital energy into NSs via tidal interaction, effectively constituting an additional loss of orbital energy, thus hastening the merging. 
To leading order, the tidal contributions of two individual stars add up linearly and the gross effect is dictated by the mutual tidal deformability \cite{hind10}, defined as $\tilde{\Lambda} = 16(m_1+12m_2)m_1^4\Lambda_1/13M^5 +(1\leftrightarrow2)$ \cite{flan08,hind08},
where $M=m_1+m_2$ is the total mass of the binary, $m_1$ and $m_2$ are the masses of two NSs, and $\Lambda_1$ and $\Lambda_2$ are their tidal deformabilities. 
The phase shift owing to this tidal expense of orbital energy has been analytically investigated to 2.5 post-Newtonian (PN) order \cite{damo12}, and phenomenologically modeled by fusing the effective-one-body (EOB) results with numerical relativity (NR) outcomes \cite{bohe17,diet17,kawa18}. Some simulations show that the overtone effects are at least one order of magnitude smaller (e.g., \cite{lack17,diet17}), which may not be discernible from the observed signal in the near future.

On the other hand, dynamic tides could also cause a dephasing that could be more dominant than the dephasing caused by equilibrium tides if a mildly spinning NS (with a spin $\nu_s\ge 400$ Hz) is involved \cite{stei21,kuan22} (see also \cite{tsok19} for spin effects dominating the equilibrium tides). It is difficult to determine the theoretical limits on the spin of neutron star members, which depends on the formation and evolution path of binaries (see \cite{lori08} for a review). In particular, some of the known BNS systems, though not coalescing ones, in the Galaxy harbour a NS with a few milliseconds period, e.g., J1807-2500B \cite{lync12} and J1946+2052 \cite{stov18}. These fast-rotating NSs are thought to be able to maintain their spin up to merger \cite{zhu18}. The significance of considering such dynamical tidal effects is highlighted by these systems, even though they may not be in the majority.
The effect of dynamical tides is dominated by the $\ell=2$ $f$-mode excitation over its overtones with harmonic number $\ell>2$. Also, the contribution of other modes, e.g., $g$-modes, is negligible due to their significantly weaker (by a few orders of magnitude) coupling to the tidal field (e.g., \cite{kokk95,shib94}).

In the following years, the sensitivity curve of the existing detectors, e.g., the upgraded LIGO \cite{abbo18} or the Voyager \cite{adhi20} may be improved by a factor 2 to 3, thus the effects of tidally excited $f$-mode should be included in the waveform modelling \cite{will22} especially the spin-induced tidal effects \cite{diet17b} (see also \cite{puec23} for this aspect with 3rd generation detectors).
In particular, tidal excitations in spinning NSs are more prominent than those in non-spinning cases because the stellar spin amplifies the tidal-induced dephasing by 
(i) reducing the frequency of $f$-modes with positive (negative) winding number if the spin is (anti-)aligned with orbital axis \cite{stei21,kuan22}, 
(ii) modifying the tidal Love number \cite{land15,pani15}, and 
(iii) manifesting high-order spin-tidal interaction (e.g., coupling between electric and magnetic tidal moments; \cite{cast22}).
Among these three effects, the spin-induced modification in modal frequencies is the most significant one since the leading corrections are linear to spin, while the other two effects are at least of quadratic order to spin. We will specify ourselves to the leading order effects, and thus the modification to (ii) and (iii) will be ignored. The quadratic-in-spin correction in the mode frequency is not taken into account as well, which will result in $10\%$ ($\gtrsim16\%$) deviation for $\nu_s=700$~Hz ($10^3$~Hz) (cf.~\cite{krug19}). In addition, we will restrict ourselves to the quadrupolar component of both kinds of tidal effects (static and dynamical) as the overtones of them are at least one order of magnitude smaller than the $\ell=2$ component \cite{schm19}. While we treat dynamical tides at the linear level (i.e., no excitation of daugther modes), it is worth pointing out a study on the influence of non-linear, dynamical tides in waveform phasing \cite{yu23}.

Within the anticipated accuracy of detectors in the near future, the effects of static and dynamical tides will be largely captured by the tidal deformability, the frequency of $f$-mode $\omega_f$, and its coupling strength to tidal field $Q_f$ \cite{kuan22}. These parameters are highly sensitive to the nuclear equation of state (EOS), and the stellar mass. Even though the masses of members in a BNS can be rather accurately estimated from the chirp mass ${\cal M}=(m_1m_2)^{3/5}/M^{1/5}$ and the symmetirc mass ratio $\eta=m_1m_2/M$ \cite{cult94,kokk94}, the uncertainty of the EOS poses difficulties in measuring the structural parameters of NSs, e.g., radius, tidal Love number, $f$-mode's frequency, etc. To extract these internal parameters, a more delicate analysis should be implemented to monitor the evolution of tidal contribution in the waveform phasing, which entails an improved treatment of tidal effects.

In this short paper, we aim to provide an approximate expression for the tidal dephasing, where the influences of spin, tilt angle, and mass ratio will be taken into account. The proposed effective tidal waveforms can explore regions of the parameter space that were previously inaccessible. It should be noted that a similar  but somewhat different quantity -- effective tidal deformability \cite{hind16,stei16} -- has been developed for slow and aligned spins \cite{stei21} and implemented to construct waveform baseline \cite{bern15}.
It is natural, that the expression depends on the EOS since the dynamical tidal response of NSs is sensitive to it. For example, spin-induced modulation in the modal frequencies depends on the EOS. Thus the accurate extraction of the tidal parameters will lead to drastic constraints on the EOS. 
To demonstrate the imprint of $f$-mode excitation instead of providing precise analysis, we make use of the PN evolution of BNSs to derive the waveform phasing $\Psi(\Oo)$ through (see, e.g., Eqs.~(14) and (17) of \cite{damo12})
\begin{align}
    \frac{d^2\Psi}{d\Oo^2}=\frac{2}{\Oo/dt}
\end{align}
for the orbital frequency $\Oo$. Although the PN formalism breaks down in the last few orbits \cite{boyl07,baio10,bern12}, the effects can be quantitatively studied since PN prescription is accurate for BNS evolution over most of the late inspiral phase \cite{droz99,buon09,radi14,ajit14,abbo16,mukh21}. In addition, the systematic error of ignoring tidal effects in non-spinning case is already comparable to the error budget between PN and hybrid (EOB+NR) waveforms \cite{read13} let alone the tidal dephasing for spinning case.

Section \ref{approx} forms the main body of this article, where we introduce the approximate baseline for the tidal dephasing with $f$-mode response taken into account. Our key result is the fitting formula for the proposed tidal form factor [Eq.~\eqref{eq:ff_aligned}], that accommodates the spin modification in both adiabatic and dynamical tidal imprints on waveforms. The influence of inclined spins (Section \ref{tilted}) and mass ratio (Section \ref{ratio}) of the binary in this formula follows with each developed into a section. Some discussion is offered in Section \ref{discussion}.

\section{Approximant of dynamical tides}\label{approx}
Taking into account the accuracy of waveform templates needed for the existing and near-future detectors, it is adequate to drop higher order terms, e.g., the spin-tidal coupling, and to adopt the simplified ansatz, $\Psi_{\text{GW}}\simeq \Psi_{\text{pp}}+\Psi_{\text{spin}}+\Psi_{\text{tidal}}$, for the GW phasing (see, e.g., Eq.~(1) of \cite{diet19}).
Here the first term, $\Psi_{\text{pp}}$, is the phasing caused by sizeless, non-spinning objects, the second term, $\Psi_{\text{spin}}$, is due to the spin effects, while the last term, $\Psi_{\text{tidal}}$, is due to tidal response of the extended objects. 

Contrary to the point-particle contributions, $\Psi_{\text{pp}}+\Psi_{\text{spin}}$, the tidal dephasing, $\Psi_{\text{tidal}}$, is becoming significant in the high frequency portion of gravitational waveform where frequency is $f_\text{gw}\gtrsim 100$ Hz (see, e.g., Fig.~2 in \cite{harr18}).
In particular, dephasing due to $f$-mode excitation largely \emph{accumulates in the last 3 seconds of coalescing binaries}.
For slowly-rotating NSs, the dephasing is characterised by $\tilde{\Lambda}$ (static tides) and it is larger than the dephasing due to mode excitations. However, for stellar spins of $\nus>500$ Hz, the dephasing caused by the excitation of $f$-mode may dominate over the influence of static tides, regardless of the EOS \cite{kuan22}.

\begin{figure}
	\centering
	\includegraphics[width=\columnwidth]{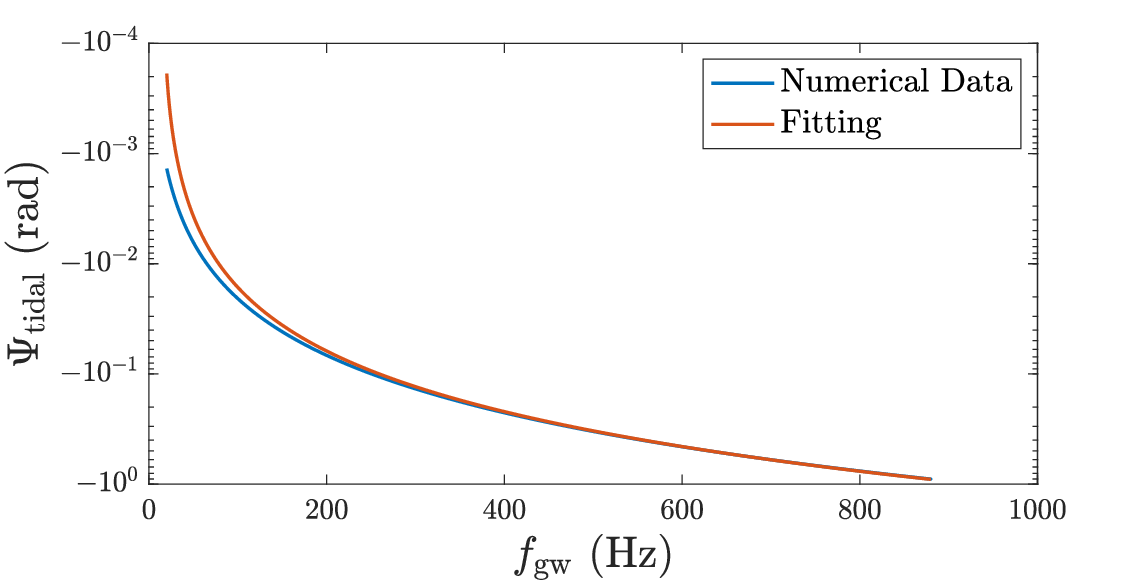}
	\caption{Tidal dephasing as functions of GW frequency, where the numerical result (blue) is approximated by the fitting formula \eqref{eq:anastz} (red) when $f_{\rm gw}\ge400$~Hz.
	}
	\label{fig:anastz}
\end{figure}

We utilize the code presented in \cite{kuan22} to construct the waveform including the leading-order $f$-mode effect. The waveform phasing generated by this code has been compared to several phenomenological and EOB templates for the non-spinning scenarios, as described in the previously mentioned paper (see Figs.~1 and 2 therein).
For equal-mass binaries, we numerically fit the tidal dephasing \emph{caused by a single NS} \footnote{The gross tidal dephasing can be obtained summing over the contribution from individual NSs.} into the form (cf.~Eq.~(31) of \cite{damo12}):
\begin{align}\label{eq:anastz}
    &\Psi_\text{tidal} = -\frac{117}{128}\left[ \Lambda\times F\left(\nus,\log\Lambda\right)\right]x^{5/2}\nonumber\\
    &\,\,\,\times\left(1+\frac{3115}{1248}x-\pi x^{3/2}+\frac{28024205}{3302208}x^2-\frac{4283}{1092}\pi x^{5/2}\right),
\end{align}
where $x=\left[\pi(m_1+m_2)f_\text{gw}\right]^{2/3}$, and we introduce a form factor, $F(\nus,\log\tilde{\Lambda})$, that depends on the NS spin and its tidal deformability. Although  we assume a constant multiplier in this short paper (see below for justification), the effective enhancement of tidal deformability by a mode excitation scales with the mode's amplitude in reality, thus varying with time \cite{hind16,kawa18}. However, the accumulation of tidal dephasing, both static and dynamic, during the early stages when the frequency of the associated gravitational wave is less than a few hundred Hz, is negligible \cite{harr18}. We, therefore, only need to ensure that the form factor can reliably reproduce the dephasing when the orbital frequency is high. In particular, we plot in Fig.~\ref{fig:anastz} the evolution (blue) together with the fitting (red) of tidal dephasing for a specific equal-mass binary with each member having $M=1.39M_{\odot}$ and pertaining to EOS APR4 \cite{apr4}. The error budget of the fitting is $\lesssim0.01$ rad, and we see that the matching between data and the fitting formula is especially satisfactory when $f_\text{gw}>400$ Hz with deviation $<2.5\%$. The fitting of tidal phase shift suggests that one can determine a biased Love number when dynamic tides are ignored in the analysis. This approach assumes the use of a point particle description and incorporates the same correction for spin-induced quadrupole moments.

\begin{figure}
	\centering
	\includegraphics[width=\columnwidth]{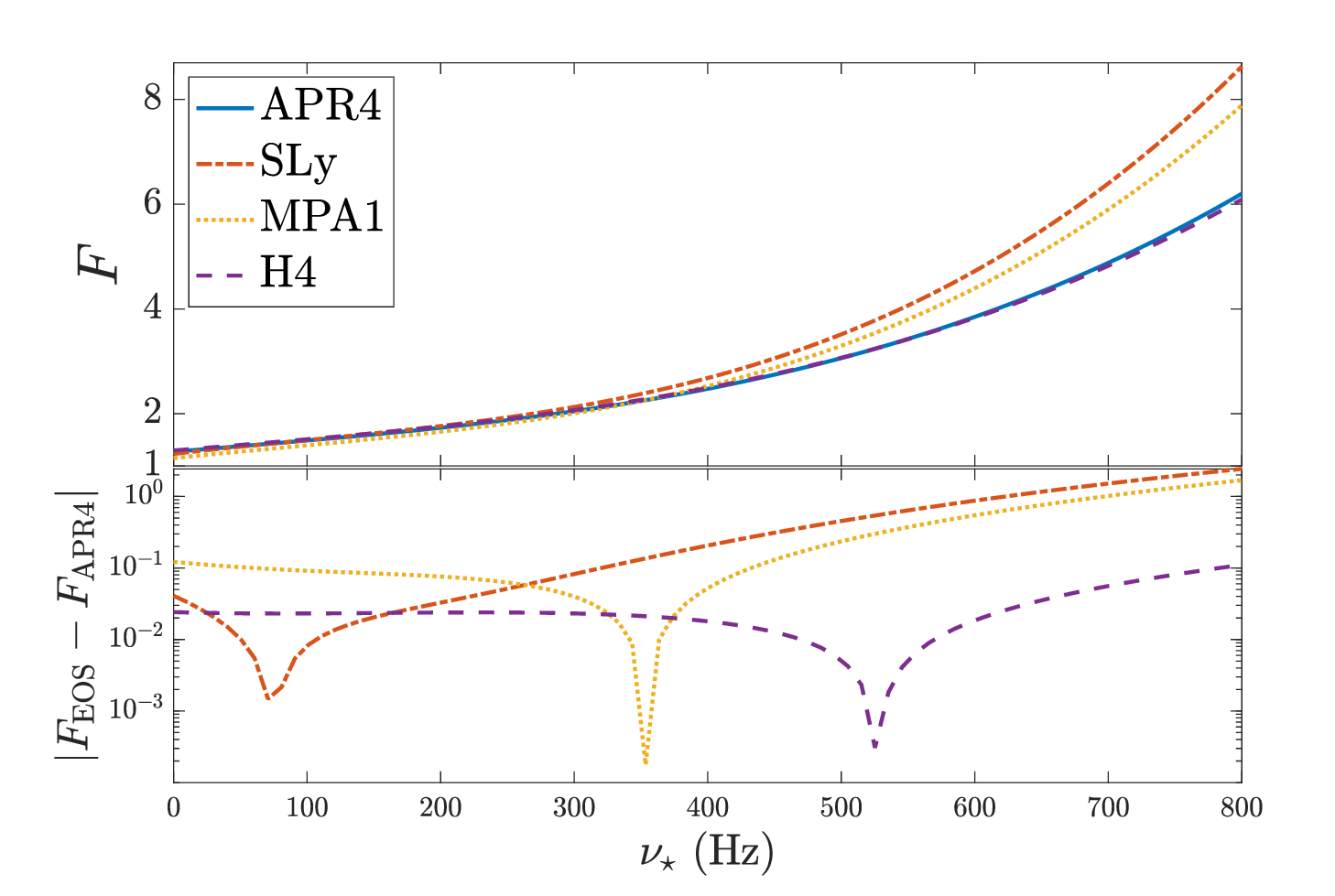}
	\caption{\emph{Top:} Form factors of NSs having tidal deformability of $\Lambda=800$ for a range of aligned spins up to $800$ Hz, and for EOS APR4 (solid), SLy (dot-dashed), MPA1 (dotted), and H4 (dashed). \emph{Bottom:} The difference between the form factor of indicated EOS from that of EOS APR4. All shown quantities are presented as functions of $\nu_\star$.
	}
	\label{fig:Fc}
\end{figure}

For equal-mass binaries with both NSs pertaining to EOS APR4, we calculate waveforms for various spins and stellar masses of equal-mass binaries, where both members follow the EOS APR4. By extracting the point-particle contribution from the resulted waveform, the tidal dephasing is obtained. We have found the following least-square fitting formula that is applicable to masses within the range  $1\,\,M_\odot$ to the maximum mass of non-spinning configurations, and spins up to 600 Hz:
\begin{align}\label{eq:ff_aligned}
    F({\tilde \nu_s}&,\log\Lambda) 
    \approx 
    \left(1.958+2.919{\tilde \nu_s} +0.094{\tilde \nu_s}^2+7.463{\tilde \nu_s}^3\right) \nonumber\\
    & -\left(1.033+1.103{\tilde \nu_s}+1.530{\tilde \nu_s}^2-0.002{\tilde \nu_s}^3\right)\log\Lambda \nonumber\\
    &+\left(0.389+0.343{\tilde \nu_s}+0.936{\tilde \nu_s}^2-0.718{\tilde \nu_s}^3\right)(\log\Lambda)^2 \nonumber\\
    &-\left(0.041-0.004{\tilde \nu_s}+0.287{\tilde \nu_s}^2-0.483{\tilde \nu_s}^3\right)(\log\Lambda)^3
\end{align}
where ${\tilde \nu_s}=\nu_s/1000\text{ Hz}$.
In reality, the exact value of each coefficient depends on the EOS, while the deviation for different EOS  becomes noticeable only for large enough $\nus$. To demonstrate how, for a varying spin, the form factor depends on the EOS, we consider some other EOSs varying in stiffness, viz.~SLy \cite{sly}, MPA1 \cite{MPA1}, and H4 \cite{H}.
The form factors of these varying-spin sequences of NSs with fixed $\Lambda$ are plotted in Fig.~\ref{fig:Fc}. We see that a noticeable difference appears when $\nus$ is larger than a few hundred Hz, e.g., a difference of $>10^{-1}$ manifests only for $\nu_\star>400$~Hz. As mentioned above, the dynamical tides are measured by two parameters $\omega_f$ and $Q_f$, and thus it may seem suspicious that these two quantities are absent in the argument of the form factor. The reason is that they are closely related to $\log\Lambda$ (e.g., Eqs.~(21)-(23) of \cite{kuan22}). 
We note that Eq.~\eqref{eq:ff_aligned} accounts for the excitations of both  $m=-2$ and $m=+2$ $f$-modes. However, for larger spins the frequencies of the $m=-2$ modes increase \cite{krug19,krug21} and hence their tidal excitation is suppressed.  Modes with winding numbers $m=\pm1,\,\,0$ are not excited if the spin is aligned with the orbit, while they will be relevant for non-trivial inclinations (see below).
Ignoring $f$-mode effects in the waveform analysis means that the form factor is practically absorbed into the tidal deformability, i.e., the effective tidal deformability ${\Lambda}F$ would be (mistakenly) viewed as the physical tidal deformability ${\Lambda}$. Overseeing the influence of the tidal form factor will thus lead to an overestimated $\tilde{\Lambda}$ (e.g., \cite{abbo19}), though the deviation may be negligible for slowly rotating NSs (see also below).

\section{Inclined neutron stars}\label{tilted}
If the spin-axis is misaligned with the orbital angular momentum, the tilt angle $\Theta$ will 
(i) lead to excitations of $f$-modes with $m\ne2$ \cite{ho99,lai06}, and 
(ii) introduce $\Theta$-dependent tidal coupling strength to each mode \cite{double}. 
We note that $Q_f$ would be modified by $\Theta$, while $\omega_f$ would not. The reduced coupling is non-linearly affecting the waveform:  the binary shrinks a bit slower for a weaker $f$-mode excitation, which, however, indicates that the mode has more time to grow. The ramification of the lower spin is similar to the effect of larger $\Theta$ since a lower spin leads also to a weaker $f$-mode excitation. Although the effects of the inclination angle and magnitude of spin in GW phasing are entangled, they can potentially be distinguished since they contribute differently to the form factor (see below and Appendix \ref{apdx}).
\begin{table}
	\centering
	\caption{The form factor, $F$, for the $f_{m=2}$\,-mode of a  member of an equal-mass BNS system. Each NS has $1.4M_{\odot}$, and pertains to the EOS APR4.
	Several spin rates and tilt angles are considered.  The columns list results for fixed spins, while the rows for fixed inclination angles. 
	}
	\begin{ruledtabular}
	\begin{tabular}{c|ccccc}
		            & 0 Hz   & 30 Hz  & 100 Hz & 400 Hz & 600 Hz \\
		\hline 
		$0^{\circ}$ & 0.5789 & 0.6056 & 0.6759 & 1.1860 & 1.9721 \\
		$10^{\circ}$ & 0.5615 & 0.5875 & 0.6556 & 1.1504 & 1.9131 \\
		$30^{\circ}$ & 0.4387 & 0.4590 & 0.5123 & 0.8991 & 1.4962 \\
		$60^{\circ}$ & 0.1832 & 0.1917 & 0.2140 & 0.3757 & 0.6261
	\end{tabular}
	\end{ruledtabular}
	\label{tab:sp_incl}
\end{table}
\begin{table}
	\centering
	\caption{Same as Tab.~\ref{tab:sp_incl}, for the  $f_{m=1}$-mode. We note that the results of aligned NSs (i.e., $\Theta=0^{\circ}$) vanish since in this case the $f_{m=1}$\,-modes are not excited. }
	\begin{ruledtabular}
	\begin{tabular}{c|ccccc}
		& 0 Hz & 30 Hz & 100 Hz & 400 Hz & 600 Hz \\
		\hline
		$5^{\circ}$ & 0.0044 & 0.0045 & 0.0047 & 0.0061 & 0.0073 \\
		$10^{\circ}$ & 0.0172 & 0.0176 & 0.0186 & 0.0238 & 0.0286 \\
		$30^{\circ}$ & 0.1260 & 0.1289 & 0.1360 & 0.1743 & 0.2100 \\
		$60^{\circ}$ & 0.2443 & 0.2498 & 0.2636 & 0.3378 & 0.4069 \\
	\end{tabular}
	\end{ruledtabular}
	\label{tab:sp_incl2}
\end{table}

{A spin $\boldsymbol{S}$ pointing towards the same hemisphere as the orbital angular momentum $\boldsymbol{L}$ (i.e.,  $\boldsymbol{S}\cdot\boldsymbol{L}>0$)} will reduce the mode frequencies with $m\ge1$, while $Q_f$ will not be affected noticeably until very large values of spin \cite{pani15,land15} ({for spins $\boldsymbol{S}\cdot\boldsymbol{L}<0$} the method can be trivially extended).
In this case the earlier onset of excitation enhances the dephasing effect \cite{kuan22}.
The form factors of these two modes, denoted as $F_{m=1}$ and $F_{m=2}$, are given in Appendix \ref{apdx} with the same ansatz as Eq.~\eqref{eq:ff_aligned} while each coefficient is expressed, in addition, as a function of tilt angle $\Theta$. 

The explicit dependence on $\Theta$ hints at that \emph{we may estimate the inclination by implementing the waveform templates including dynamical tides to the match filtering method.} We note that this inclination effect is encoded in the tidal effects, which is already minute. The information of inclination can also be sifted from early stages of insprialling by filtering the detected data stream against waveforms that takes into account orbital precession in the context of PN \cite{apos94, kidd95} and/or EOB \cite{osso20,khal23,ramo23,pomp23} treatments. Although the later approach is more promising avenue to probe the tilted spin, we note that the measurability depends on the orientation of the binary compared to the observer. It is, on other hand, insensitive to the relative orientation in the former aspect.
Here we provide an assessment on the significance of $\Theta$ in waveforms when dynamical tides are taken into account. In Tabs.~\ref{tab:sp_incl} and \ref{tab:sp_incl2}, we collate the dephasing generated by, respectively, $m=2$ and $m=1$ $f$-mode excitations for a NS with $M=1.4M_{\odot}$ member of an equal-mass binary. We see that the excitation of $m=1$ mode contributes more to the dephasing than the $m=2$ mode when $\Theta\gtrsim60^{\circ}$ (the relative strength of the excitation of the two modes is encoded in the Wigner D-functions; see, e.g., Sec.~2.4 of \cite{double} and the references therein). However, when the effects of $m=1$ and $m=2$ are combined, the form factor (i.e., $F=F_{m=1}+F_{m=2}$) is reduced for increasing $\Theta$. For example, there is a $\sim24\%$ drop in $F$ when the tilt angle is increased from $\Theta=30^{\circ}$ to $\Theta=60^{\circ}$.

\section{Mass ratio}\label{ratio}
As far as the leading-order effects are concerned, the individual contribution to the tidal dephasing can be added linearly, implying that the {``mutual form factor''} can be defined via 
\begin{align}
    \tilde{F} =\frac{F_1+\beta F_2}{1+\beta}
\end{align}
for $\beta=q^4(12+q)/(1+12q)$ and $q=m_2/m_1$, where $F_i$ is the form factor of star $i$.
In this way we can combine $\tilde{F}$ with the ``mutual Love number'' $\tilde{\Lambda}$ as follows:
\begin{align}\label{eq:mutual_F}
 {\tilde \Lambda}_{\rm comb}\equiv\tilde{\Lambda}\cdot \tilde{F} = \frac{16}{13}\frac{(m_1+12m_2)m_1^4}{M^5}\Lambda_1 F_1+(1\leftrightarrow2)\, .
\end{align}
The tidal deformability, ${\tilde \Lambda}_{\rm comb}$ (Eq.~\ref{eq:mutual_F}), is measured from the GW analysis, while for estimating the ``true tidal deformability'' ${\tilde \Lambda}$ one needs to divide it by a factor of $\tilde{F}$. Therefore, the favoured EOS would be softer than expected if the $f$-mode effects are ignored. In the following, we take the specific event GW170817\footnote{The nature of this event cannot really be distinguished by jointly considering the GW and electromagnetic emissions, as suggested in \cite{hind19}. We assume it was a BNS event in the present work however.} as an example to illustrate how the inclusion of $f$-mode effects would change the estimate on $\tilde{\Lambda}$ and thus the constraints on the EOS.

In the analysis carried out by the LIGO-VIRGO collaboration, the progenitor binary consists of NSs with masses of $\sim1.48M_{\odot}$ ($\sim 1.63M_{\odot}$) and $\sim1.26M_{\odot}$ ($\sim1.18M_{\odot}$) if slow (high) spin prior is assumed \cite{abbo17,abbo19}. 
{If we assume that one of the NSs is non-spinning, and consider aligned spins of $\nus=30$ Hz and $\nus=400$ Hz as two representative cases}, we list in the 5th column of Tab.~\ref{tab:gw170817} the mutual form factor $\tilde{F}$ for each scenario. 
The effective spin, defined by 
\begin{align}
    \chi_{\rm eff}=\frac{m_1\chi_1+m_2\chi_2}{M},
\end{align}
for each case is provided in the third column, where $\chi_1$ and $\chi_2$ are the dimensionless spin parameter for star 1 and 2, respectively. This parameter for GW170817 has been constrained to $\lesssim0.1$ with $90\%$ credible interval (cf.~Fig.~6 of \cite{abbo19}), and the considered high spin cases are marginal to this constraint.
By taking the heavier (lighter) star as star 1 (2) with spin $\nu_1$ ($\nu_2$), we have the inequality $F_1>F_2$, and it can be derived from the definition of $\tilde{F}$ that $\tilde{F} \ge F_2$. In accordance, the true value of $\tilde{\Lambda}$ should be modified by at least a factor of $F_2$ from the observed tidal deformability $ {\tilde \Lambda}_{\rm comb}$. Taking the measured tidal deformability of GW170817 (i.e., ${\tilde \Lambda}_{\rm comb}\le 800$; \cite{abbo17,abbo19}), we list the true tidal deformability $\tilde{\Lambda}$ in the last column of Tab.~\ref{tab:gw170817}.

\begin{table}
	\centering
	\caption{A worked example for GW170817.
    The 1st and 2nd columns show, respectively, the (aligned) spins in Hz, and the masses for both NSs in solar mass. The 3rd column offers the effective spin.
	In the 4th column, the associated form factors of the individual stars are listed, while the 5th column provides the mutual form factor. The final column presents the estimations of $\tilde{\Lambda}$  assuming that the  $f$-mode excitation is taken into account. For the NS models used here we assumed the APR4.}
	\begin{ruledtabular}
	\begin{tabular}{cccccc}
		$\nu_1$, $\nu_2$ & $M_1$, $M_2$ & $\chi_{\rm eff}$ & $F_1$, $F_2$ & $\tilde{F}$ & $\tilde{\Lambda}$ \\
		\hline
		(\,\,\,\,\,30, \,\,0) & (1.48, 1.26) & 0.009 & (1.19, 1.20) & 1.19 & $\le$669.51 \\
		(\,\,\,\,\,\,0, \,\,30) & (1.48, 1.26) & 0.008 & (1.14, 1.26) & 1.21 & $\le$659.85 \\
		(400, \,\,\,\,0) & (1.63, 1.18) & 0.118 & (2.26, 1.23) & 1.98 & $\le$403.06 \\
		(\,\,\,\,0, 400) & (1.63, 1.18) & 0.115 & (1.12, 2.58) & 2.19 & $\le$365.06
	\end{tabular}
	\end{ruledtabular}
	\label{tab:gw170817}
\end{table}

In addition to the upper bound on $\tilde{\Lambda}$, the post-merger, electromagnetic counterpart can arguably set a lower bound since smaller $\tilde{\Lambda}$ renders a more compact remnant thus less ejecta can be scattered away during and soon after the merger \cite{radi17}, even though some investigation suggests otherwise, e.g., \cite{kiuc19,nich21}. If the electromagnetic counterpart could provide reliable constraints on $\tilde{\Lambda}$, it would supplement the bounds derived by gravitational wave data analysis.
For example, a lower bound of  $\tilde{\Lambda}$, e.g. $\tilde{\Lambda}\agt400$, will exclude the scenario that the 2nd star will be aligned and rapidly-spinning (cf.~the last row of Tab.~\ref{tab:gw170817}). However, this constraint can be lifted even for small inclinations, e.g. $\Theta=20^{\circ}$, since the value of the form  factor will be reduced  to $\tilde{F}=$1.98, leading to degenerate results. For example, the same outcome will  be derived if  the 1st star   spins at 400 Hz with trivial tilt angle while the 2nd star  is non-rotating (cf.~the last two rows of Tab.~\ref{tab:gw170817}).

\section{Discussion}\label{discussion}
In this short paper, we put tidal dephasing into perspective by proposing a close form, though as a fitting formula, for the tidal form factor $F$ [Eq.~\eqref{eq:ff_aligned}].
The form factor is $\ge1$, indicating an enhancement of tidal dephasing due to dynamical tides. In fact, the form factor presents the ratio of phase shift caused by static and dynamic tides; denoting the dephasing by the static tide as $\Psi_\Lambda$, and that by $f$-mode excitation as $\Psi_f$, these two quantities are related via $\Psi_f = (F-1) \Psi_\Lambda$.
However, we note that the above relation holds only for waveforms having a cutoff frequency $\ge400$~Hz (cf.~Fig.~\ref{fig:anastz}).

Owing to the dependence of the mode excitation on the tidal deformability $\Lambda$, the spin $\nus$, inclination $\Theta$, and the EOS, the coefficients of the analytic expression for the form factor are functions of the aforementioned parameters (see Appendix \ref{apdx}). Generally speaking, the form factor will be larger for increasing spin since the excitation of the $f$-mode kicks in earlier, and thus absorbs more orbital energy. In certain cases, we see that the tidal form factor can reach up to $F\sim2$ (Tab.~\ref{tab:sp_incl}), implying that neglecting dynamical tides, i.e., forcing the mutual form factor of the binary to $\tilde{F}=1$, may lead to an overestimate of $\tilde{\Lambda}$ by a factor of 2 [cf.~Eq.~\eqref{eq:mutual_F}]. Such a dramatic change will critically affect the {identification} of the EOS. For example, considering aligned spins, the reduced upper bound on $\tilde{\Lambda}$ for GW170817 would favor softer EOS (cf.~Tab.~\ref{tab:gw170817}). 
Furthermore, an overestimation of $\tilde{\Lambda}$ will affect the derivation of the (effective) compactness of the long-lived NS remnant \cite{mano21}, else will result in some inconsistencies between the measurements of these two quantities. 

In the future, with the most sensitive interferometers, the detailed waveform during merger and post-merger phases may become detectable, providing information for the peak frequency of the merger and the remnant's  $f$-mode frequency, if a prompt collapse to a BH is staved off. 
The constraint jointly set by all these observations can only be meaningful if $\tilde{\Lambda}$ can be correctly estimated, especially for constraining EOS candidates \cite{sama18}. For example, given that the relation between $\tilde{\Lambda}$ and ${\cal M}$ is sensitive to the EOS (see, e.g., Fig.~8 of \cite{kawa18}; also \cite{zhao18}) and that the latter quantity can be determined with great precision in waveform analysis, the uncertainty of constraining EOS is thus set mainly by the accuracy of the estimate of $\tilde{\Lambda}$.
On the other hand, a tilt angle leads to smaller form factors (Tabs.~\ref{tab:sp_incl} and \ref{tab:sp_incl2}) as a result of weakened $f$-mode excitation. The reduction in the form factor is sensitive to $\Theta$, and thus may provide a novel hope for estimating the inclination, while we note that the effects of the tilt angles of both stars tangle together. Any further effort toward this direction would be beneficialby supplementing the current effective treatment of precession \cite{schm12,schm15} as well as the future analytic efforts.

It is important to note that, here we do not take into account the spin-induced correction in the tidal deformability (or the correction in the tidal overlap), which, however, is not anticipated to influence the effect studied here. The extent to which the $f$-mode is excited, can roughly be evaluated by the product of its coupling strength to the tidal field and the reciprocal of its frequency, viz.~$\eta \approx \omega_f^{-1} Q_f$, since the growth rate of the amplitude scales with $Q_f$ and the duration of the excitation is inversely proportional to the mode frequency $\omega_f$. The expansion in terms of spin of  this ``efficiency factor'' $\eta$, reads $\eta = \eta_0 + a_1\nus\eta_1 + a_2\nus^2\eta_2 + O(\nus^3)$.
Here, $\eta_0$ is the efficiency of $f$-mode excitation in non-spinning NS, and $a_1$ and $a_2$ are some EOS-dependent constants. In this expansion, only the mode frequency modulation will contribute to $a_1$, while the spin-induced modification in $Q_f$ appears only in $a_2$ and higher-order terms. We therefore believe that the form factors obtained in the present article will not be affected in a noticeable way by the input of realistic tidal overlap in rotating NSs.

Although there are numerical methods aimed at modeling tidal effects in the waveform by matching the simulated signal to EOB  results (e.g., \cite{hoto15,hoto16,dudi18,kawa18,diet19}, the initial data (ID) for numerical relativity (NR) simulations is constructed in a way that does not involve mode excitation. This means that the dynamic response of neutron stars (NSs) is not resolved. The ID is designed to represent an equilibrium state of the binary system less than 100 ms before the merger, during which the excitation of the $f$-mode may not be negligible. As a result, the constructed ID may not accurately represent the binary system's state in reality, especially when compared to an EOB waveform that includes significant tidal excitations (as mentioned in \cite{gamb23}). However, this limitation is relatively minor for slowly-spinning binaries, as the excitation of $f$-modes is weak in such cases.

\section*{Acknowledgement}
Data generated from numerical codes are reported in the body of the paper. Additional data will be made available upon reasonable request. We would like to express our gratitude to the anonymous referees for their valuable feedback, which enhances the quality of our current work.


\bibliographystyle{apsrev4-2}
\bibliography{references}

\appendix
\section{Form factors}\label{apdx}

The form factors for $f_{m=2}$ and $f_{m=1}$ -modes can be approximated by
\begin{align}
    &F_{m=2}(\Theta,\nuk,\log\Lambda)= 
    (a_1+a_2\nuk+a_3\nuk^2+a_4\nuk^3)\nonumber\\
    &\quad\quad
    +(a_5+a_6\nuk+a_7\nuk^2+a_{8}\nuk^3)\log\Lambda\nonumber\\
    &\quad\quad
    +(a_{9}+a_{10}\nuk+a_{11}\nuk^2+a_{12}\nuk^3)(\log\Lambda)^2\nonumber\\
    &\quad\quad
    +(a_{13}+a_{14}\nuk+a_{15}\nuk^2+a_{16}\nuk^3)(\log\Lambda)^3 \\
   \nonumber\\
    &F_{m=1}(\Theta,\nuk,\log\Lambda)= (b_1+b_2\nuk+b_3\nuk^2+b_4\nuk^3)\nonumber\\
    &\quad\quad+(b_5+b_6\nuk+b_7\nuk^2+b_8\nuk^3)\log\Lambda\nonumber\\
    &\quad\quad+(b_9+b_{10}\nuk+b_{11}\nuk^2+b_{12}\nuk^3)(\log\Lambda)^2\nonumber\\
    &\quad\quad+(b_{13}+b_{14}\nuk+b_{15}\nuk^2+b_{16}\nuk^3)(\log\Lambda)^3,
\end{align}
where ${\tilde \nu}=\nu/(1\,{\rm kHz})$. 

Considering spins up to 600 Hz, and tilt angles up to $\Theta=110^{\circ}$ for the fitting formula, we find the coefficients are given as
\begin{align}
    &a_1 = 0.975+0.213\IC-4.795\IC^2+5.377\IC^3-1.743\IC^4,\nonumber\\
    &a_2 = 1.453+0.326\IC-7.225\IC^2+8.138\IC^3-2.652\IC^4,\nonumber\\
    &a_3 = 0.044+0.061\IC-0.164\IC^2+0.065\IC^3,\nonumber\\
    &a_4 = 3.718+0.839\IC-18.884\IC^2+21.559\IC^3-7.131\IC^4,\nonumber\\
    &a_5 = -0.514-0.113\IC+2.528\IC^2-2.833\IC^3+0.918\IC^4,\nonumber\\
    &a_6 = -0.55-0.129\IC+2.902\IC^2-3.378\IC^3+1.141\IC^4,\nonumber\\
    &a_7 = -0.760-0.141\IC+2.604\IC^2-2.137\IC^3+0.405\IC^4,\nonumber\\
    &a_8 = -0.004-0.012\IC+1.713\IC^2-3.178\IC^3+1.491\IC^4,\nonumber\\
    &a_9 = 0.194+0.042\IC-0.950\IC^2+1.064\IC^3-0.344\IC^4,\nonumber\\
    &a_{10} = 0.171+0.040\IC-0.997\IC^2+1.228\IC^3-0.439\IC^4,\nonumber\\
    &a_{11} = 0.465+0.101\IC-1.379\IC^2+0.861\IC^3-0.029\IC^4,\nonumber\\
    &a_{12} = -0.355-0.093\IC+0.378\IC^2+0.666\IC^3-0.614\IC^4,\nonumber\\
    &a_{13} = -0.021-0.005\IC+0.101\IC^2-0.113\IC^3+0.037\IC^4,\nonumber\\
    &a_{14} = 0.002+0.001\IC+0.029\IC^2-0.061\IC^3+0.031\IC^4,\nonumber\\
    &a_{15} = -0.143-0.035\IC+0.480\IC^2-0.359\IC^3+0.051\IC^4,\nonumber\\
    &a_{16} = 0.24+0.062\IC-0.841\IC^2+0.656\IC^3-0.108\IC^4
\end{align}
and
\begin{align}
    &b_1 = 0.010-0.304\IC +5.682\IC^2-9.056\IC^3+3.825\IC^4,\nonumber\\
    &b_2=0.010-0.210\IC+3.783\IC^2-6.009\IC^3+2.533\IC^4,\nonumber\\
    &b_3=-0.076\IC+1.631\IC^2-2.660\IC^3+1.148\IC^4,\nonumber\\
    &b_4=0.004-0.075\IC+1.333\IC^2-2.069\IC^3+0.847\IC^4,\nonumber\\
    &b_5=-0.005+0.159\IC-2.980\IC^2+4.751\IC^3-2.007\IC^4,\nonumber\\
    &b_6=-0.005+0.107\IC-1.749\IC^2+2.754\IC^3-1.154\IC^4,\nonumber\\
    &b_7=0.003+0.025\IC-0.772\IC^2+1.329\IC^3-0.603\IC^4,\nonumber\\
    &b_8=-0.003-0.001\IC-0.116\IC^2+0.110\IC^3,\nonumber\\
    &b_9=0.002-0.059\IC+1.118\IC^2-1.782\IC^3+0.753\IC^4,\nonumber\\
    &b_{10}=0.002-0.041\IC+0.640\IC^2-1.002\IC^3+0.419\IC^4,\nonumber\\
    &b_{11}=-0.002-0.009\IC+0.303\IC^2-0.532\IC^3+0.246\IC^4,\nonumber\\
    &b_{12}=0.002+0.003\IC-0.025\IC^2+0.081\IC^3-0.058\IC^4,\nonumber\\
    &b_{13}=0.006\IC-0.118\IC^2+0.188\IC^3-0.079\IC^4,\nonumber\\
    &b_{14}=0.004\IC-0.054\IC^2+0.083\IC^3-0.034\IC^4,\nonumber\\
    &b_{15}=0.001\IC-0.030\IC^2+0.055\IC^3-0.027\IC^4,\nonumber\\
    &b_{16}=-0.003\IC+0.042\IC^2-0.074\IC^3+0.035\IC^4,
\end{align}
where EOS APR4 is assumed and ${\tilde \Theta}=\Theta/100^{\circ}$.

\end{document}